\documentclass[a4paper,fleqn,usenatbib]{mnras}

\usepackage{graphicx}	
\usepackage{amsmath}	
\usepackage{amssymb}	

\title[The Nele asteroid family]{On the age of the Nele asteroid family}
\author[V. Carruba, D. Vokrouhlick\'{y}, D. Nesvorn\'{y} and
 S. Aljbaae]{V. Carruba$^{1,3}$\thanks{E-mail: vcarruba@gmail.com},
             D. Vokrouhlick\'{y}$^{2}$,
             D. Nesvorn\'{y}$^{3}$ and
             S. Aljbaae$^{1}$\\
$^{1}$S\~{a}o Paulo State University (UNESP), School of Natural Sciences
 and Engineering, Guaratinguet\'{a}, SP, 12516-410, Brazil \\
$^{2}$Institute of Astronomy, Charles University, V Hole\v{s}ovi\v{c}k\'{a}ch
 2, Prague 8, CZ-18000, Czech Republic\\
$^{3}$Department of Space Studies, Southwest Research Institute, Boulder, 
 CO, 80302, USA}

\date{Accepted 2018 March 15.  Received 2018 March 13; in original form 2018 February 8.}
\pubyear{2018}

\begin{document}
\label{firstpage}
\pagerange{\pageref{firstpage}--\pageref{lastpage}} 
\maketitle

\begin{abstract}
 The Nele group, formerly known as the Iannini family, is one of the
 youngest asteroid families in the main belt. Previously, it has been
 noted that the pericenter longitudes $\varpi$ and nodal longitudes
 $\Omega$ of its largest member asteroids are clustered at the present
 time, therefore suggesting that the collisional breakup of parent body must
 have happened recently. Here we verify this conclusion by
 detailed orbit-propagation of a synthetic Nele family and show that the
 current level of clustering of secular angles of the largest Nele family
 members requires an approximate age limit of $4.5$~Myr. Additionally, we
 make use of an updated and largely extended Nele membership to obtain,
 for the first time, an age estimate of this family using the Backward
 Integration Method (BIM). Convergence of the secular angles in a purely
 gravitational model and in a model including the non-gravitational forces
 caused by the Yarkovsky effect are both compatible with an age younger than
 $7$~Myr. More accurate determination of the Nele family age would require
 additional data about the spin state of its members.

\end{abstract}

%

\section{Introduction}
\label{sec: intro}

Asteroid families form as a result of collisions among asteroids. The orbits of 
fragments originating from each break-up event subsequently evolve due to 
gravitational and non-gravitational effects such as planetary perturbations and 
the Yarkovsky effect \citep{Bottke_2002, Vokrouhlicky_2015}.  The present
orbital distribution of family members is therefore a combination of the
initial spread produced by the original ejection velocity field and later
dynamical evolution. Separating the two effects is often difficult, because
the velocity field is a priory unknown and because the strength of the
Yarkovsky effect depends on various parameters, such as the asteroid density
and surface thermal conductivity, that are poorly constrained. 

Young asteroid families (i.e., families that formed less than $\sim 20$~Myr 
ago), are useful in this context, because of our ability to precisely model
the dynamics of asteroids on short timescales. For example, in the case of 
a young asteroid family it is often possible to demonstrate the past
convergence of the pericenter longitudes $\varpi$ and nodal longitudes
$\Omega$. Such a convergence is expected because the ejection velocities are
typically only a small perturbation of the orbital speed. All fragments
should therefore have about the same values of $\varpi$'s and $\Omega$'s
immediately after the parent body breakup, which becomes apparent when the
orbits of young family members are integrated backward in time from the present
epoch \citep[e.g.,][]{Nesvorny_2002, Nesvorny_2003}. 

Modeling young families provides valuable information about their age 
and membership. It may also help to constrain key parameters of the Yarkovsky
effect, because the Yarkovsky effect needs to be taken into account to
precisely reconstruct the past convergence of angles
\citep[e.g.,][]{Nesvorny_2004}.  In some cases, such as for the Karin cluster,
it was even shown that strict convergence criteria allow for a detection of
the radiation torque known as the YORP effect \citep{Carruba_2016a}.

In this work, we focus our attention on the case of the Nele cluster
\citep[Family Identification number, or FIN, 520 in][]{Nesvorny_2015}. This 
family, previously also known as the Iannini family, was suggested to be younger
than $5$~Myr in \citet{Nesvorny_2003}, because the $\varpi$'s and $\Omega$'s of 
its members were found to be clustered even at the present epoch. Using
a larger dataset of numbered and multi-opposition asteroids for which the
proper orbital elements were determined, \citet{Milani_2014} recently found
that (1547) Nele is likely the largest member in the former Iannini cluster
and thus opted to rename the cluster after this body. We adopt this point of
view, but readers should be aware of the previous name used for this group,
if necessary. Here we take advantage of a still larger data-set of asteroids
for which proper orbital elements were determined and made available at the
AstDyS website in June 2017 to extend and improve the analysis of both
\citet{Nesvorny_2003} and \citet{Milani_2014}. More importantly, using 
methods also developed in \citet{Carruba_2016a, Carruba_2017} for the Karin 
and Veritas families, we demonstrate the past convergence of 
secular angles of asteroids in the Nele family, and obtain an age
estimate of this family using the direct backward orbit integration of
its members.

\section{Family identification and dynamical properties}
\label{sec: fam_ide}

Two sets of membership lists are currently available for the Nele family.
\citet{Nesvorny_2015} used the Hierarchical Clustering Method 
\citep[HCM,][]{Bendjoya_2002} and a cutoff of $25$~m/s to obtain a family
of $150$ members, which is $\simeq 8$ times more than the $18$ members 
available at the time of \citet{Nesvorny_2003}. This family was still called
Iannini in \citet{Nesvorny_2015}. Preference to this name was based on
\citet{Nesvorny_2003}, who observed that (1547) Nele is offset in proper 
eccentricity and inclination with respect to location of other large
asteroids in this cluster. However, when smaller members were later
associated with this cluster in larger databases, this was no longer the
case. For that reason \citet{Milani_2014} proposed to call this cluster
the Nele family. Indeed, more recent and quite larger datasets
of proper orbital elements that become available at the AstDyS
website in June~2017 \citep[http://hamilton.dm.unipi.it/astdys][]{Knezevic_2003}
confirm this trend. Automated methods of family identification at the
AstDyS site have resulted in the detection of $344$ members of the Nele family.
Observing this evolution, we also adopt the Nele attribution for this
family name and basically use the membership available at the AstDyS 
website, performing the following brief analysis.

The family is very compact and isolated. Using the criteria of 
\citet{Carruba_2016}, we identified objects in the local background of this
group. Specifically, we used the synthetic proper elements from the 
AstDyS site, and selected orbits with values between 
0.2663 to 0.2710 in proper $e$ and between 0.2103 to 0.2130 in proper $\sin{i}$ 
(four standard deviations of the Nele family distribution around the family 
barycenter). Values of proper $a$ were chosen from 2.638~au to 2.651~au
(the family range $\pm0.02$~au). Only $81$ additional background objects were
found using this method (atop to the $344$ identified Nele members). These
additional asteroids were found to be randomly scattered in the selected
box of proper orbital elements showing no tendency of clustering.
This indicates that these background asteroids are, most likely,
not family members  and we will not include them in the following analysis.
Using HCM for objects in this local background we found that 92.7\% of the
425 asteroids in the region join the Nele family for cutoff values of 
$50$~m/s. All asteroids become members of the family for cutoff values larger 
than $75$~m/s.

\begin{figure*}
  \centering
  \begin{minipage}[c]{0.49\textwidth}
    \centering \includegraphics[width=3.1in]{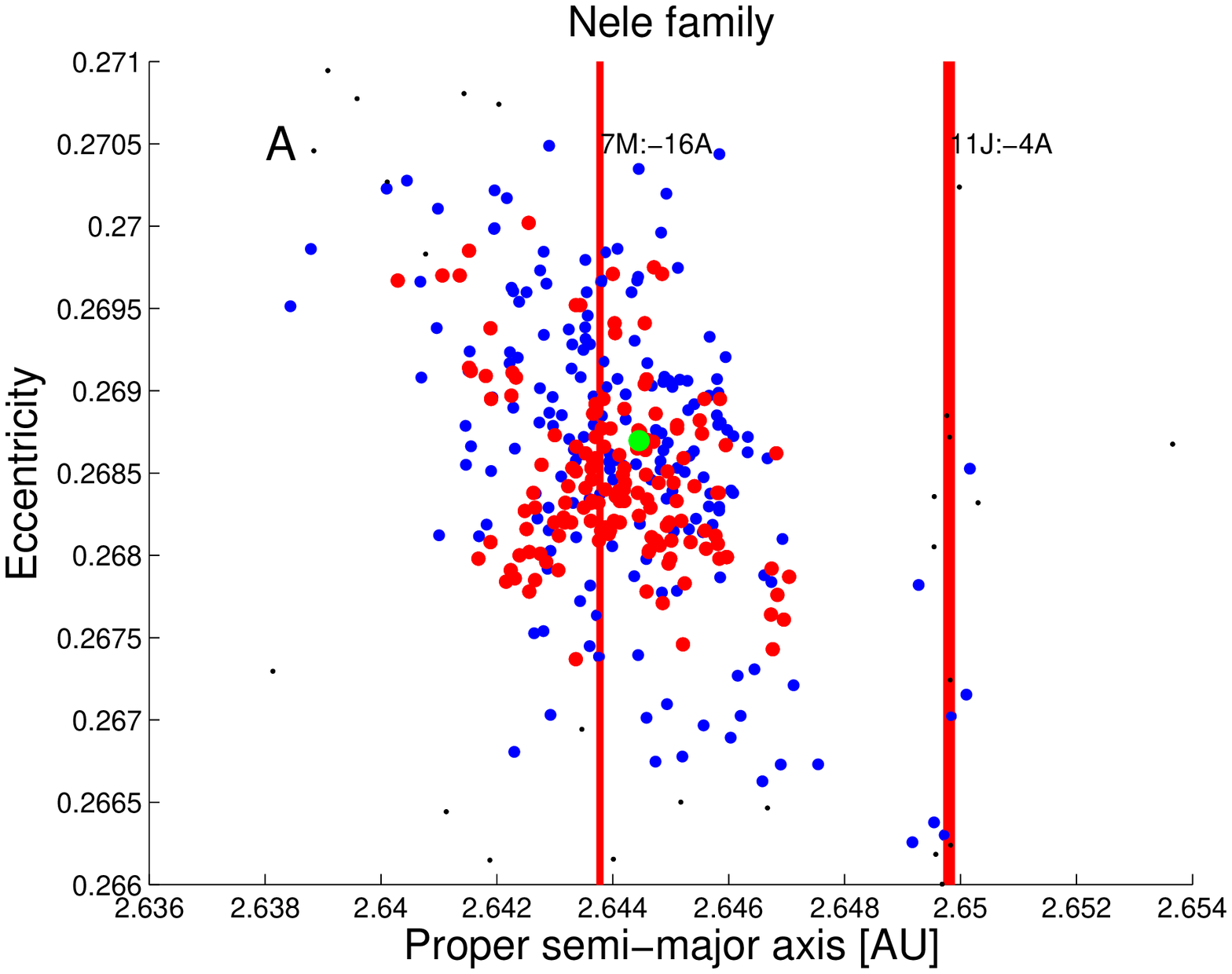}
  \end{minipage}%
  \begin{minipage}[c]{0.49\textwidth}
    \centering \includegraphics[width=3.1in]{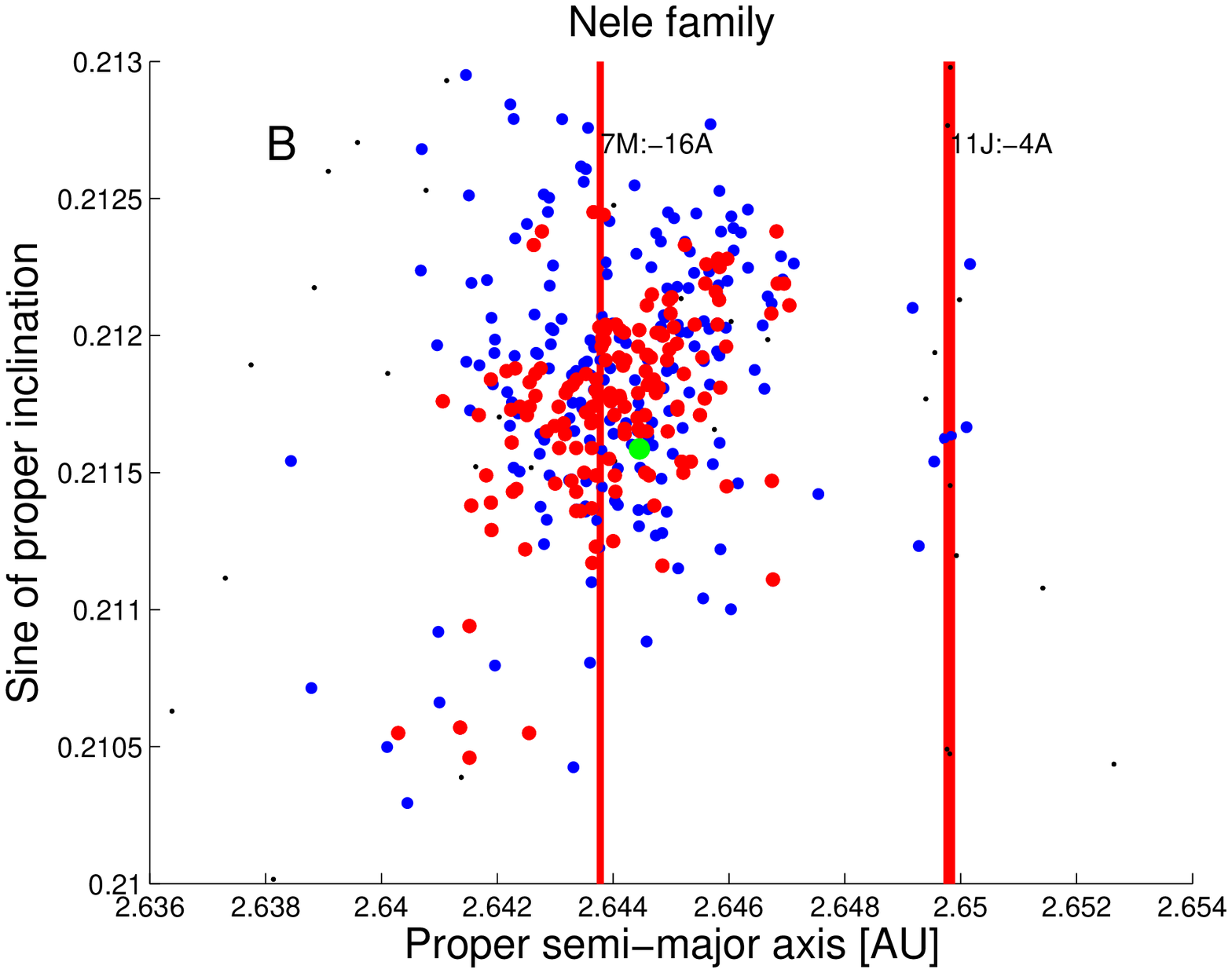}
  \end{minipage}
\caption{Proper elements, $(a,e)$ in panel A and $(a,\sin{i})$ in panel B,
 of Nele family members (blue dots) and background asteroids (black dots).
 For comparison, the red dots display the orbital location of the Iannini
 asteroid family, as available from \citet{Nesvorny_2015}. The large
 full green dot shows the orbital location of (1547) Nele itself. The vertical
 lines display the location of the principal mean motion resonances in
 the region.}
\label{fig: iannini_back}
\end{figure*}

The \citet{Milani_2014} and \citet{Nesvorny_2015} families are shown
in Fig.~\ref{fig: iannini_back}. We identified
all secular resonances up to order six in the Nele family
region, but no member of the family was found in a secular resonance with 
planets or massive asteroids. Secular dynamics therefore seems 
unimportant. \citet{Gallardo_2014} identified the 11J-4A resonance with 
Jupiter at $a\simeq 2.6498$~au and the 7M-16A resonance with Mars
at $a\simeq 2.6438$~au as the two main mean motion resonances in the region.
The 11J-4A affects seven family members at large-end of the semi-major 
axis values, but most Nele objects are not significantly affected by either
of the resonances.

\begin{figure}
\centering
\centering \includegraphics [width=0.45\textwidth]{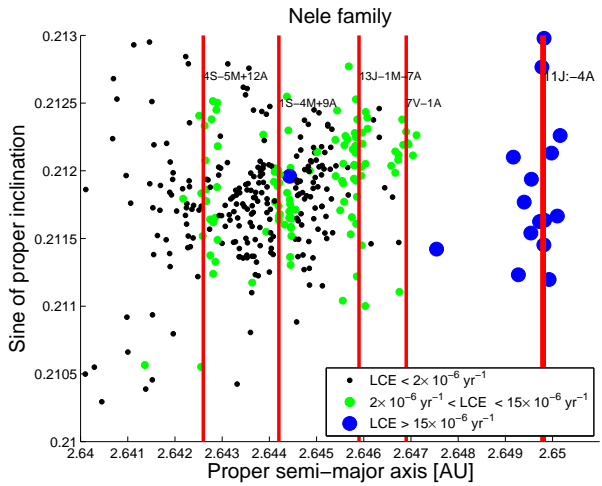}
\caption{A $(a,\sin{i})$ projection of asteroids in the local background of
 the Nele family. Objects with Lyapunov times $T_{\rm L}$ longer than
 $5\times 10^5$ yr are shown as black dots (these are the most stable
 orbits). Objects with $6.7 \times 10^4 < T_{\rm L}  < 5 \times 10^5$~yr are
 displayed as green full circles, while the objects with $T_{\rm L} < 6.7
 \times 10^4$~yr are shown as blue full circles (for most part affected by
 the 11J-4A mean motion resonance).}
\label{fig: Lyap_ae}
\end{figure}

To better understand the role of chaotic dynamics, we identified asteroids
in different ranges of the Lyapunov Characteristics Exponents (LCE).
The LCE estimates were obtained from AstDyS. Fig.~\ref{fig: Lyap_ae}
shows objects with LCEs in three different intervals (the figure caption
specifies the equivalent limits in terms of the Lyapunov times $T_{\rm L}$, 
defined as the inverse of LCEs). Apart from the chaotic region associated with 
the 11J-4A resonance already discussed, three mildly chaotic areas were
also found: one at $\simeq 2.6426$~au, possibly associated with the 4S-5M+12A
three-body resonance, one at $\simeq 2.6444$~au, possibly related to the 
1S-4M+9A and four-body resonances such as 15J-1S-1M+3A, and one at
$\simeq 2.6464$~au, which is probably related to the 13J-1M-7A three-body
resonance and the 7V-1A two-body resonance with Venus
\citep[e.g.,][]{Nesvorny_1998, Gallardo_2014}. 
No chaotic region was found at the semi-major axis of the 7M-16A resonance. 
Both (1547) Nele and (4652) Iannini, the family largest bodies, have mildly 
chaotic orbits (i.e., $6.7 \times 10^4 < T_{\rm L}  < 5 \times 10^5$~yr). 

\section{Physical properties}
\label{sec: phys_prop}

\citet{Nesvorny_2015} listed Iannini as an S-type family, with a mean geometric
albedo $p_V$ of 0.32. This analysis is confirmed for the new larger Nele
family found by \citet{Milani_2014} and recent updates at the AstDyS, that
identify this group as an S-type family of a mean albedo 0.355. There is 
very limited information available about other physical properties of the 
members of the Nele family. Only 7 objects have photometric data in the Sloan 
Digital Sky Survey-Moving Object Catalog data \citep[SDSS-MOC4;][]{Ivezic_2001} 
from which the taxonomic information can be obtained using the method of
\citet{DeMeo_2013}, and only 14 objects have geometric albedo and absolute
magnitude information available in the WISE and NEOWISE, AKARI, or IRAS
databases \citep{Mainzer_2016, Masiero_2012, Ishihara_2010, Ryan_2010}.

Of the seven objects with taxonomic information based on broad-band SDSS data, 
4 belong to the S-complex
(4652 Iannini, K-type, 78908, Q-type, 146444, S-type, and 161328, S-type), and
three to the C-complex (1547 Nele, TD-type, and 151032 and 164726, both
C-types).  These data do not allow us to reach any conclusion on the
taxonomy of the Nele family. Geometric albedo data is more helpful. Of the
14 objects with WISE albedo, only two, (185380) and (43766) have an albedo
lower than 0.2. The median value of the geometric albedo is of 0.315, and the
mean value is equal to 0.293, essentially confirming the analysis of
\citet{Nesvorny_2015} and \citet{Milani_2014}. Based on this data, the Nele
family is most likely an S-complex family, and the asteroids 151032,
164726, and 185380 are either potential interlopers or objects for which the
Sloan-derived taxonomy is misleading. Interestingly, (1547) Nele was classified 
as a TD object in a Tholen taxonomy based on two color indexes, but has an
albedo of $0.20$, barely compatible with that of other family members.

\begin{figure}
\centering
\centering \includegraphics [width=0.45\textwidth]{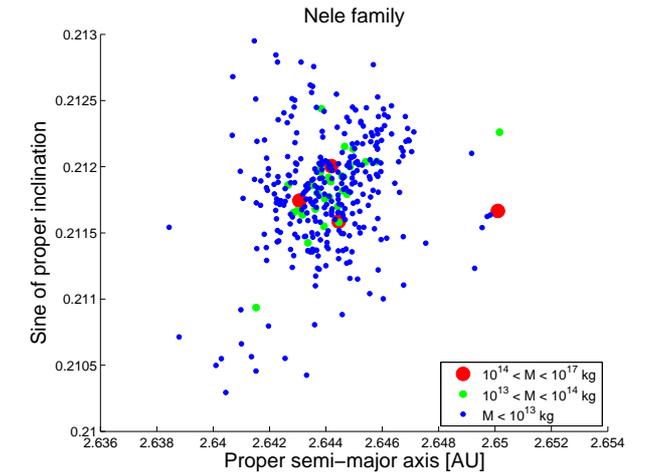}
\caption{An $(a,\sin{i})$ projection of asteroids near the orbital location of 
 the Nele family. The color and size of the symbols reflect the estimated 
 asteroid mass. Asteroid (43766) 1988 CR4, represented by the large red
 symbol at $a\simeq 2.65$~au away from the center of the Nele family and
 in the 11J-4A resonance, is suggested to be an interloper from the
 background population of objects.}
\label{fig: masses}
\end{figure}

To better understand the distribution of largest members of the family, we
estimated the masses assuming objects to be spherical with a bulk density
equal to 2500 kg m$^{-3}$ (typical value of S-type objects; this is close to 
the guessed value of 2275 kg m$^{-3}$ reported at the AstDyS site). For 
objects with available WISE albedo data, we used the WISE $p_V$ value to
estimate their radius from the absolute magnitude (Eq.~1 as in
\citet{Carruba_2003}). For all other objects, we used $p_V = 0.315$, which
is the median value of this family.  Fig.~\ref{fig: masses} shows the
results. Only three objects have estimated masses larger than $10^{14}$ kg
and diameters $D>5$~km and these are (1547) Nele ($D\simeq 17.3$~km),
(4652) Iannini ($D \simeq 5.3$~km), and (43766) 1988 CR4
($D \simeq 7.2$~km). However, the fact that (43766) 1988 CR4 is an
albedo outlier of this family and its location in the 11J-4A mean-motion
resonance away from other large asteroids in the family both suggest this
body is an interloper in the cluster. (81550) 2000 HU23 is the only other
object with an estimated mass larger than $10^{14}$~kg, and has a
diameter of $\simeq 4.9$~km.

\section{Age estimates of the Nele family}
\label{sec: p_age_est}

Recently, \citet{Spoto_2015} estimated the age of the Nele family
using the method of V-shape fitting in the domain of semi-major axis
and inverse diameters. In this method the distribution of asteroids
is binned for different values of asteroid inverse diameters, and the
asteroids with minimum and maximum semi-major axis values are selected.
The bin sizes are chosen so that the differences in the number of members
in two consecutive bins never exceeds one standard deviation. Errors
are assigned on the asteroids inverse diameters, and outliers are rejected
using an automatic outliers rejection scheme. The slopes of the two sides
of the family V-shape are then computed, with their errors. Ages and their
uncertainty are then estimated using a Yarkovsky calibration for the values
of the semimajor axis drift rate $da/dt$, based on the radar results for 
asteroid (101955) Bennu \citep[][and subsequently rescaled to the family
heliocentric location and assumed bulk densities of the members]{Chesley_2014}.

\begin{figure}
\centering
\centering \includegraphics [width=0.45\textwidth]{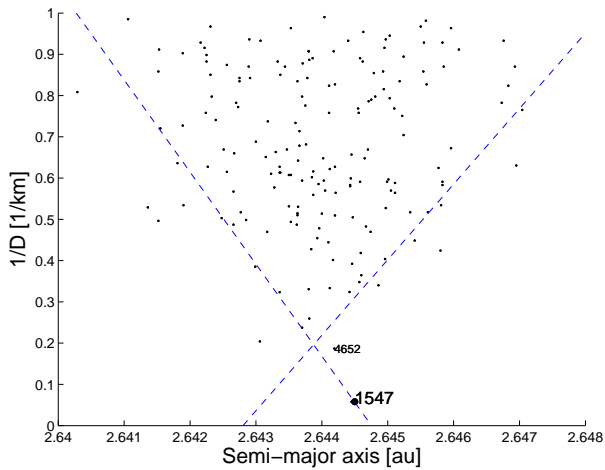}
\caption{The V-shape in the $(a,1/D)$ domain for the Nele family. The black
 full dot displays the location of (1547) Nele, largest member in the family.}
\label{fig: V-shape}
\end{figure}

\citet{Spoto_2015} obtained values of the $1/S$ slope equal to
$-0.005\pm0.0008$ for the IN inverse slope and of $0.005\pm0.002$
for the OUT inverse slope. Once the Yarkovsky calibrations and their
errors were accounted for, this corresponds to ages of $14\pm5$ and
$15\pm7$~Myr, respectively. However, observing the suggested age limit of
$\simeq 5$~Myr from clustering of secular angles in \citet{Nesvorny_2003},
\citet{Spoto_2015} pointed out that the effects of the initial velocity 
field could be rather important for this very young family
(implying that possibly the initial spread of Nele asteroids dominates
their distribution in the V-shape domain rather than the Yarkovsky effect,
typical for older families).

In order to check these conclusions with more members available,
we repeated \citet{Spoto_2015} analysis for the Nele family with a
similar method%
\footnote{Instead of the algorithm used for automatic rejection method
 used by these authors, we used built-in numerical tools available in
 Matlab.}.
Our results are in good agreement, within the uncertainties, with the 
solution published by \citet{Spoto_2015}. Fig.~\ref{fig: V-shape} displays 
the V-shape for the solution using the family center,%
\footnote{Another solution using (1547) Nele instead is available on
 the AstDyS as blue curves, but values of the slopes were not available
 for this solution} 
and the range of $1/D$ values used for the size binning of the family. 
We found inverse values of the slopes of $-0.0049\pm0.0013$ for the IN 
slope and of $0.0055\pm0.0023$ for the OUT slope.

Taking now an opposite extreme assumption about the spread of Nele family
within the bounds of V-lines, we can use the data to obtain an estimate 
of the initial ejection velocity field. Assuming the initial ejection 
velocity field was isotropic (at least for small members), the available
slopes of the delimiting lines in the $(a,1/D)$ domain can inform us about the
transverse component of the ejection velocity field $V_{\rm T}$. In particular,
if $V_{\rm T}\propto 1/D^{\alpha_{\rm EJ}}$ with some exponent ${\alpha}_{\rm EJ}$,
the relationship between the displacement of a particular family member in 
proper $a$ from the center $a_{\rm c}$, $|a-a_{\rm c}|$, reads
\citep[e.g.,][]{Vokrouhlicky_2006a,Vokrouhlicky_2006b}
\begin{equation}
 \left|a-a_{\rm c}\right| = \frac{2}{n}\,V_{\rm EJ}
  \left(\frac{D_0}{D}\right)^{{\alpha}_{\rm EJ}}\cos{\theta},
\label{eq: daD}
\end{equation}
\noindent
where $n$ is the asteroid mean-motion, $V_{\rm EJ}$ is a parameter
that describes the width of the fragment ejection velocity distribution
\citep{Vokrouhlicky_2006a,Vokrouhlicky_2006b}, $D_0$ a reference size value
from magnitude-size relationship equal to $1329$ km, and $\theta$ is 
the angle of the fragment ejection velocity relative to the transverse 
direction of the parent body's orbit. The limits of the V-shape region
then correspond to $\cos{\theta}=\pm 1$. Data for the Karin and Koronis 
\citep{Nesvorny_2002,Carruba_2016} families suggest that
${\alpha}_{EV} \simeq 1$ \citep[see also][]{Bolin_2018}. Therefore, if we
assume that the V-shape for the Nele family in the $(a,1/D)$ domain observed in
Sect.~\ref{sec: p_age_est} is mostly caused by the initial ejection
velocity field, the value of the ejection parameter $V_{\rm EJ}$ would 
be related to the absolute values of the IN and OUT slopes $1/S$ by 
the relationship

\begin{equation}
  V_{\rm EJ} = \frac{n}{2|S|}.
  \label{eq: VEJS}
\end{equation}

Using the values of $a_{\rm c}$, $S_{\rm IN}$ and $S_{\rm OUT}$ from this section,
then $V_{\rm EJ}$ would be $17.2$ and $18.4$ m~s$^{-1}$, respectively. Therefore,
this approximate analysis would yield a value of
$V_{\rm EJ}=17.8\pm 0.6$ m~s$^{-1}$, which is similar to the
estimated escape velocity from (1547)~Nele itself, that is of the order of
$10$ m~s$^{-1}$ \citep[e.g.,][]{Broz_2013}. Therefore, it is conceivable that
at least 2/3 of the asteroid distribution within the V-shape limits is due to
the initial velocity field and only 1/3 is the additional component due to
the Yarkovsky spreading. With that, both \citet{Spoto_2015} and our age
estimates mentioned above would shrink to $\simeq 5$~Myr.

\begin{figure}
\centering
\centering \includegraphics [width=0.45\textwidth]{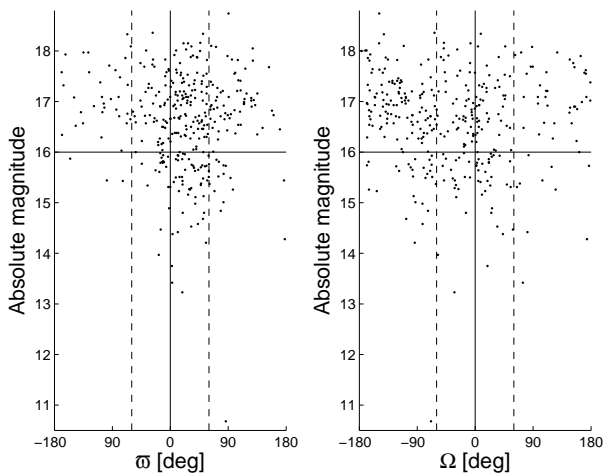}
\caption{The current $(\varpi,H)$ and $(\Omega,H)$ distributions of members of
 the Nele family, where $\varpi$ and $\Omega$ are osculating values at
 the MJD epoch 57200 and $H$ are absolute magnitudes from AstDyS site.
 The vertical black lines display $\Omega$ and $\varpi$
 equal to $0^{\circ}$, while the vertical dashed line show the $\pm 60^{\circ}$
 levels. The horizontal line displays the $H=16$ level.}
\label{fig: sec_angles_dist}
\end{figure}

Now, we come back to the idea of \citet{Nesvorny_2003} who suggested an
age limit of this cluster from today's distribution of the secular
angles of the largest members. Our plan is to carry out a more detailed
look into the problem and also check how the secular angles in the
Nele family are distributed as a function of their size (now available
with much larger dataset of its members).

Fig.~\ref{fig: sec_angles_dist} displays the current $(\Omega,H)$ and
$(\varpi,H)$ distributions for longitude of node $\Omega$ and longitude
of pericenter $\varpi$. The $\varpi$ distribution, and, to a 
lesser extent, also the $\Omega$ distribution, show tendency to cluster 
near $0^{\circ}$ for members with $H\leq 16$. Smaller members have the
secular angles dispersed uniformly in between $0^\circ$ and $360^\circ$.
This is expected, because their larger semimajor axis drift $da/dt$ due
to the Yarkovsky effect implies an additional contribution in
how the secular angles disperse, and, also, smaller fragments tend to
be ejected with higher ejection speed, further away from the family
center than larger family members, and would therefore experience
larger differences in the precession rates of the secular angles with
respect to the parent body than, in general, larger objects.
The sample of large members in the family may, therefore, be considered
to test the age limit envisaged by \citet{Nesvorny_2003}.

In order to determine the expected pace at which the secular angles of the
orbits in the Nele family diverge, we conducted the following simple
numerical experiment. At the current epoch, MJD 57200, we constructed
a synthetic cluster of 107 objects around the orbit of (1547) Nele
(this corresponds to the $H<16$ sample, Fig.~\ref{fig: sec_angles_dist}).
We used a very conservative velocity ejection speed of $3$ m~s$^{-1}$ 
for each of the fragments and assumed they were launched isotropically
in space. This renders an initial spread in both longitude of node $\Omega$ 
and pericenter $\varpi$ of only a fraction of a degree. We intentionally 
keep the initial velocity small in order not to push the age constraint for
the cluster too low.

\begin{figure*}
 \centering
 \begin{minipage}[c]{0.49\textwidth}
  \centering \includegraphics[width=3.1in]{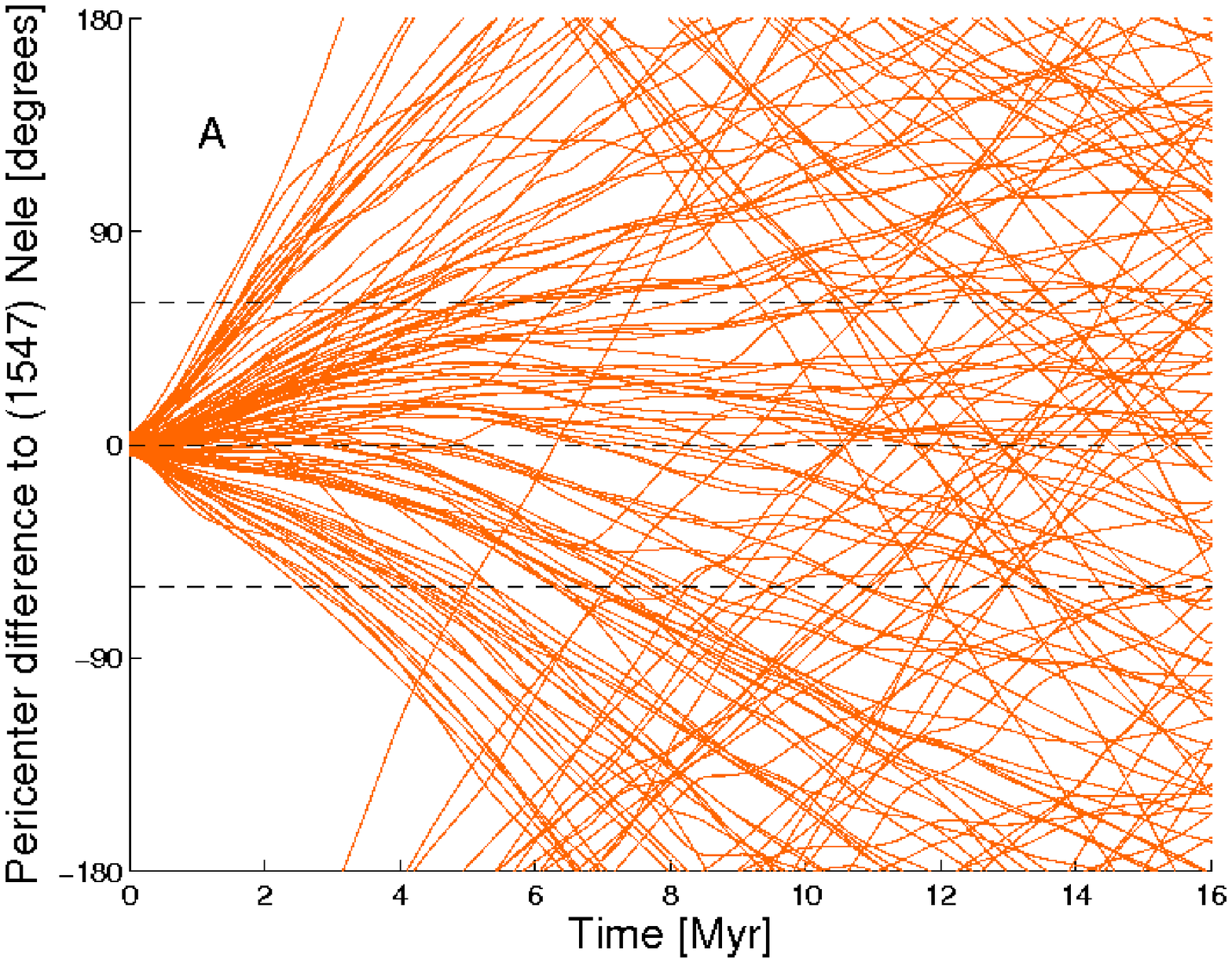}
  \end{minipage}%
 \begin{minipage}[c]{0.49\textwidth}
  \centering \includegraphics[width=3.1in]{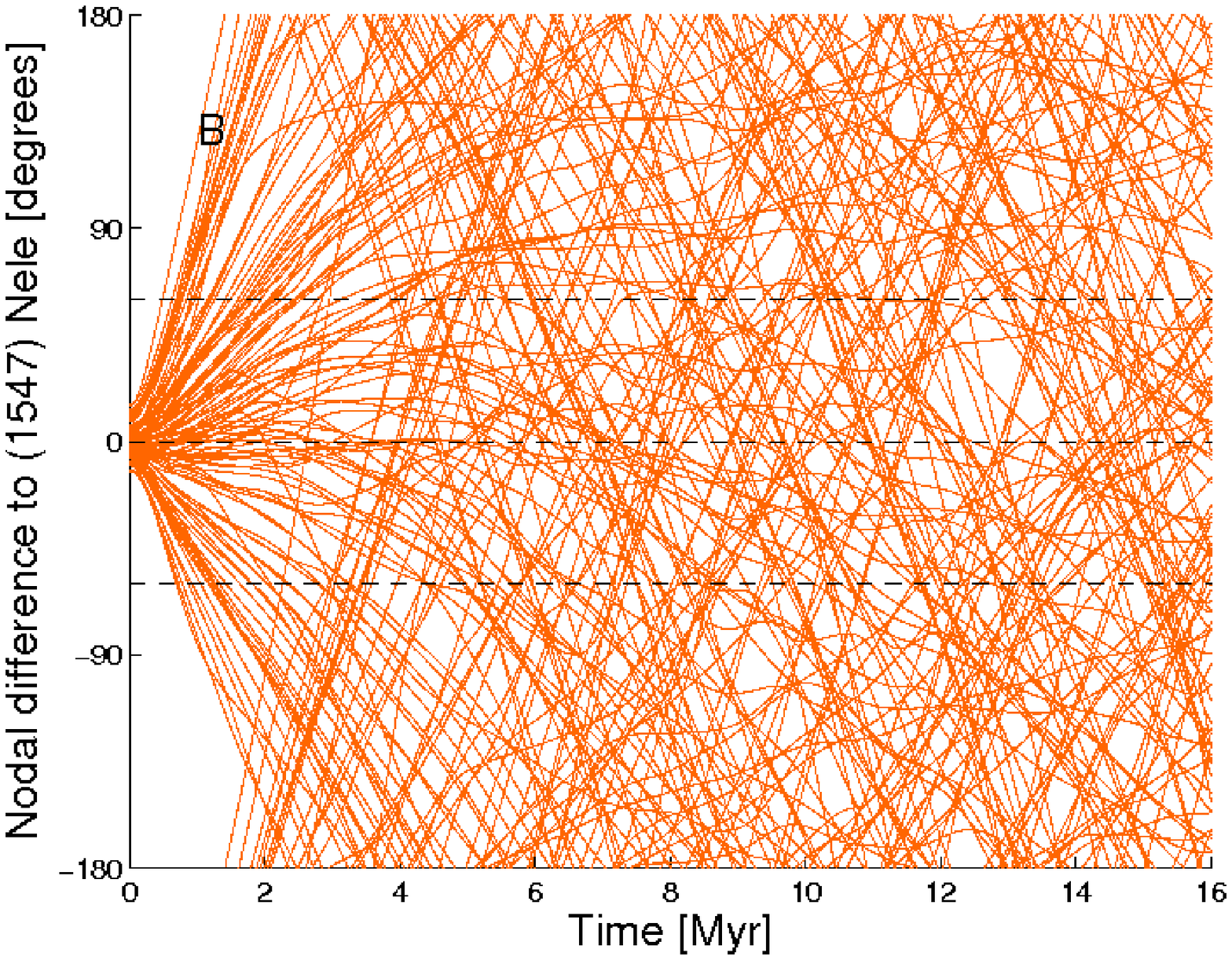}
 \end{minipage}
 \caption{The future evolution of the digitally filtered pericenter 
  (panel A) and nodal (panel B) longitudes of 107 Nele members
  of a synthetic family reproducing conditions after break-up, when
  purely gravitational forces are considered. The dashed lines show
  $0^\circ$ and $\pm 60^\circ$, for reference.}
\label{fig: Future_conv}
\end{figure*}

We numerically integrated the orbits of these objects and computed the 
filtered values of their difference in $\varpi$ and $\Omega$ with 
respect to (1547) Nele, with the same procedure discussed in 
Sect.~\ref{sec: long_conv}. The simulation included only gravitational
perturbations from the planets, driving circulation of the secular angles,
but we neglected the effects of thermal accelerations at this stage.
Fig.~\ref{fig: Future_conv} 
shows the time behavior of these differences in $\varpi$ (panel A) and 
$\Omega$ (panel B) as a function of time. As expected, the distribution
of $\Delta\varpi$ and $\Delta\Omega$ starts initially very compact
about zero and quickly spreads to become uniform in the whole range
of $0^\circ$ and $360^\circ$ in a few Myrs. While the mean rate of the
dispersion change in $\varpi$ and $\Omega$ may be estimated analytically,
the numerical approach we adopt here is more accurate (note, for instance,
that many of the trajectories in Fig.~\ref{fig: Future_conv} are not
straight lines but instead show complex dependence on time, indicating
thus chaotic evolution of the orbits).

\begin{figure}
\centering
\centering \includegraphics [width=0.45\textwidth]{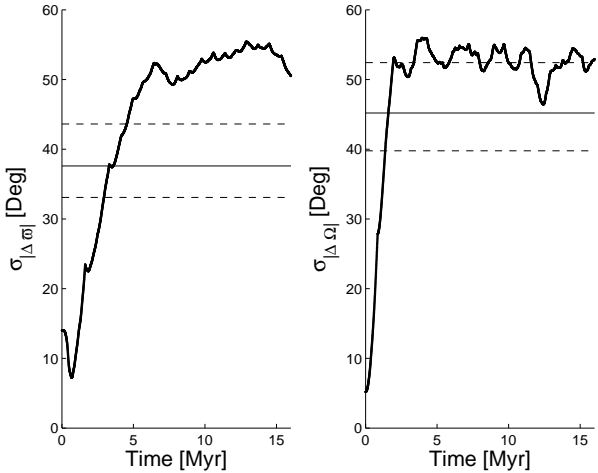}
\caption{Time behavior of the standard deviation of the absolute values 
 of differences $\Delta \varpi$ and $\Delta \Omega$ in secular angles
 with respect to (1547)~Nele. Data from numerical propagation of
 107 members of our synthetic realization of the Nele family started as
 a compact cluster with $3$ m~s$^{-1}$ ejection field of fragments.
 The horizontal black line displays the current value of the standard
 deviation for the real Nele $H < 16$ members, while the horizontal
 dashed lines show the confidence level for these values (obtained using
 methods discussed in \citet{Sheskin_2003} for a sample of $\simeq 100$
 bodies).}
\label{fig: std_nele}
\end{figure}

In order to quantify the secular angles dispersion, we computed the 
standard deviation of absolute values $|\Delta \varpi|$ and 
$|\Delta \Omega|$ distributions as a function of time for the simulated 
family. Fig.~\ref{fig: std_nele} displays our results. The horizontal 
black line indicates the current values of these standard deviations
as observed in the sample $H\leq 16$ members of the Nele
family (see Fig.~\ref{fig: sec_angles_dist}), while the dashed black
lines show the confidence level for the standard deviations of a 
population of $\simeq 100$ asteroids \citep[$0.88\, {\rm SD} < {\rm SD} <
1.16\, {\rm SD}$, with SD being the nominal value of the standard \
deviation; e.g.,][]{Sheskin_2003}. Values of the standard deviations 
for an uniform distribution of absolute values are about $51.5^{\circ}$ 
\citep[e.g.,][]{Vokrouhlicky_2006b}%
\footnote{Note that we evaluate angular separation of two angles in our
 case which is by definition limited to $0^{\circ}$ to $180^{\circ}$.
 If we were to consider uniform distribution of only one angle
 in between $0^{\circ}$ to $360^{\circ}$, the limiting standard
 deviation would be $\simeq 103^\circ$ as mentioned in 
 \citet{Vokrouhlicky_2006b}.}.
The results in $\Delta \Omega$ are not conclusive to within the errors.
This is because the divergence in $\Omega$ turns out to be quite fast
in the zone of Nele family.  Mean values of the frequency
  gradients $\frac{ds}{da}$ for
  Nele family members are equal to $-101.3\pm 0.3$~arcsec yr$^{-1}$, which
  is higher than what found for the more well-behaved Karin family,
  for which $\frac{ds}{da} = -70.0 \pm 0.2$ arcsec yr$^{-1}$
  \citep{Nesvorny_2004}.
Indeed, the data in Fig.~\ref{fig: sec_angles_dist}
confirm that nodes of even the large bodies in the Nele family are
quite spread. Luckily, data in $\varpi$ are more conclusive. The
currently observed dispersion $\Delta \varpi\simeq 37.6^\circ$ is
attained in our simulation in $3.6^{+0.9}_{-0.7}$~Myr (see panel~A in
Fig.~\ref{fig: Future_conv}), and, indeed, values of
$\frac{dg}{da}$ are relatively small for asteroids in the Nele family
($83.5\pm 0.4$ compared to $94.3\pm 0.6$~arcsec yr$^{-1}$ for Karin
asteroids).
Overall, results of this experiment indicate
that the Nele family should indeed be younger of $4.5$~Myr, strengthening
slightly the suggested limit in \citet{Nesvorny_2003}.

\section{Past convergence of the secular angles}
\label{sec: long_conv}

In this section, we attempt to improve the preliminary age estimates from
above by using a direct numerical integration of orbits for Nele members
backward in time and seeking their convergence. Following the approach 
described in \citet{Nesvorny_2003, Nesvorny_2004, Carruba_2016a} for the 
Veritas and Karin families, we first verified the ability to track
the convergence of secular angles $\varpi$ and $\Omega$ in the past
for $342$ Nele members (we excluded two albedo interlopers) 
by performing a numerical simulation where we only included
gravitational perturbations from
all planets, Mercury through Neptune, and neglected thermal accelerations
known as the Yarkovsky effect. These are expected to become more important
for small members in the family. We used the state-of-the-art symplectic
integrator known as ${\tt SWIFT\_MVSF}$ \citep[e.g.,][]{Levison_1994} with
a $10$~day timestep (verifying using a smaller sample of orbits that 
identical results are obtained for shorter timesteps too). The 
code was modified by \citet{Broz_1999} to include an online filtering of 
the osculating orbital elements.  This function allowed us to obtain both
osculating and mean orbital elements at the output from the code. 

\begin{figure*}
 \centering
 \begin{minipage}[c]{0.49\textwidth}
   \centering \includegraphics[width=3.1in]{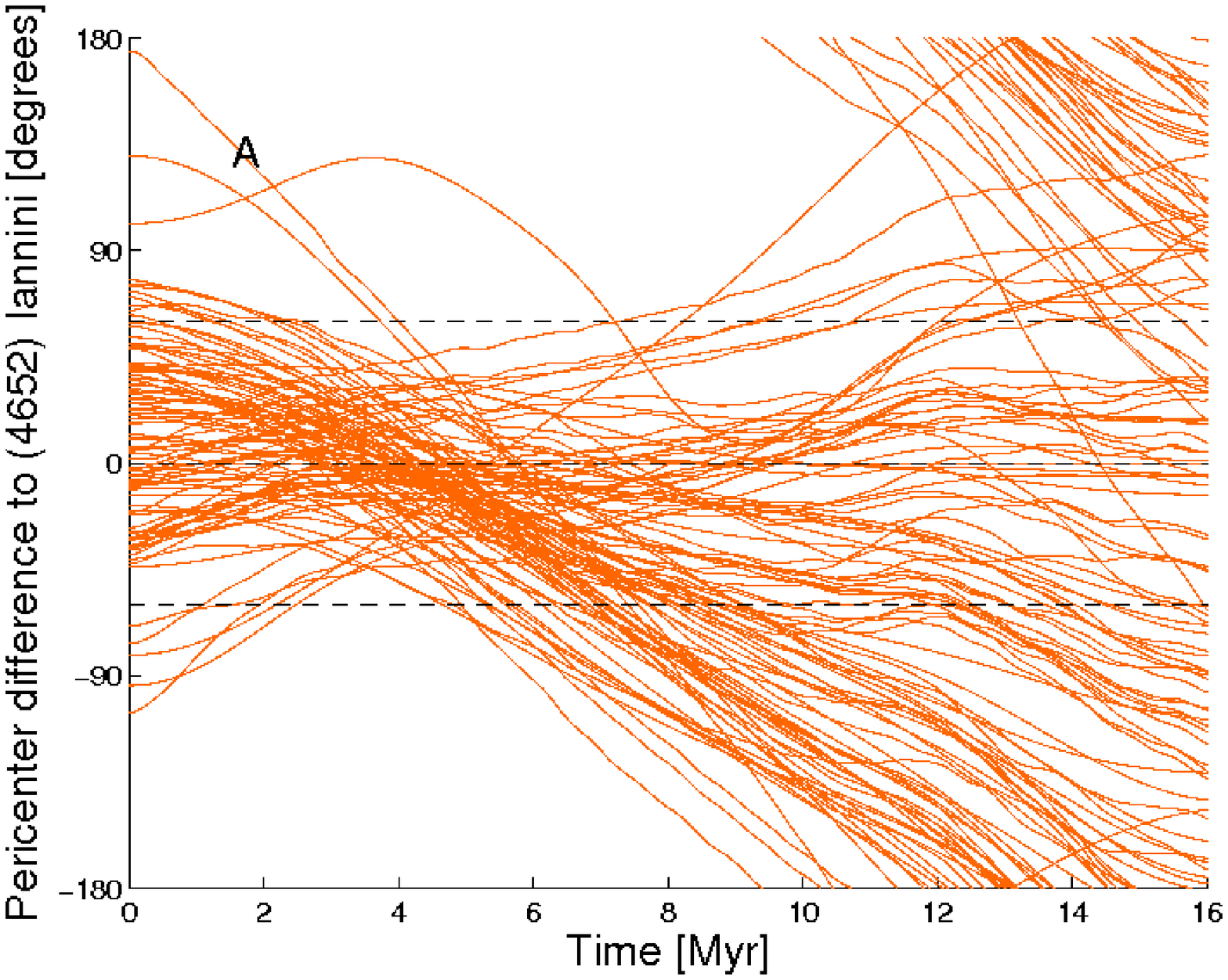}
 \end{minipage}%
 \begin{minipage}[c]{0.49\textwidth}
   \centering \includegraphics[width=3.1in]{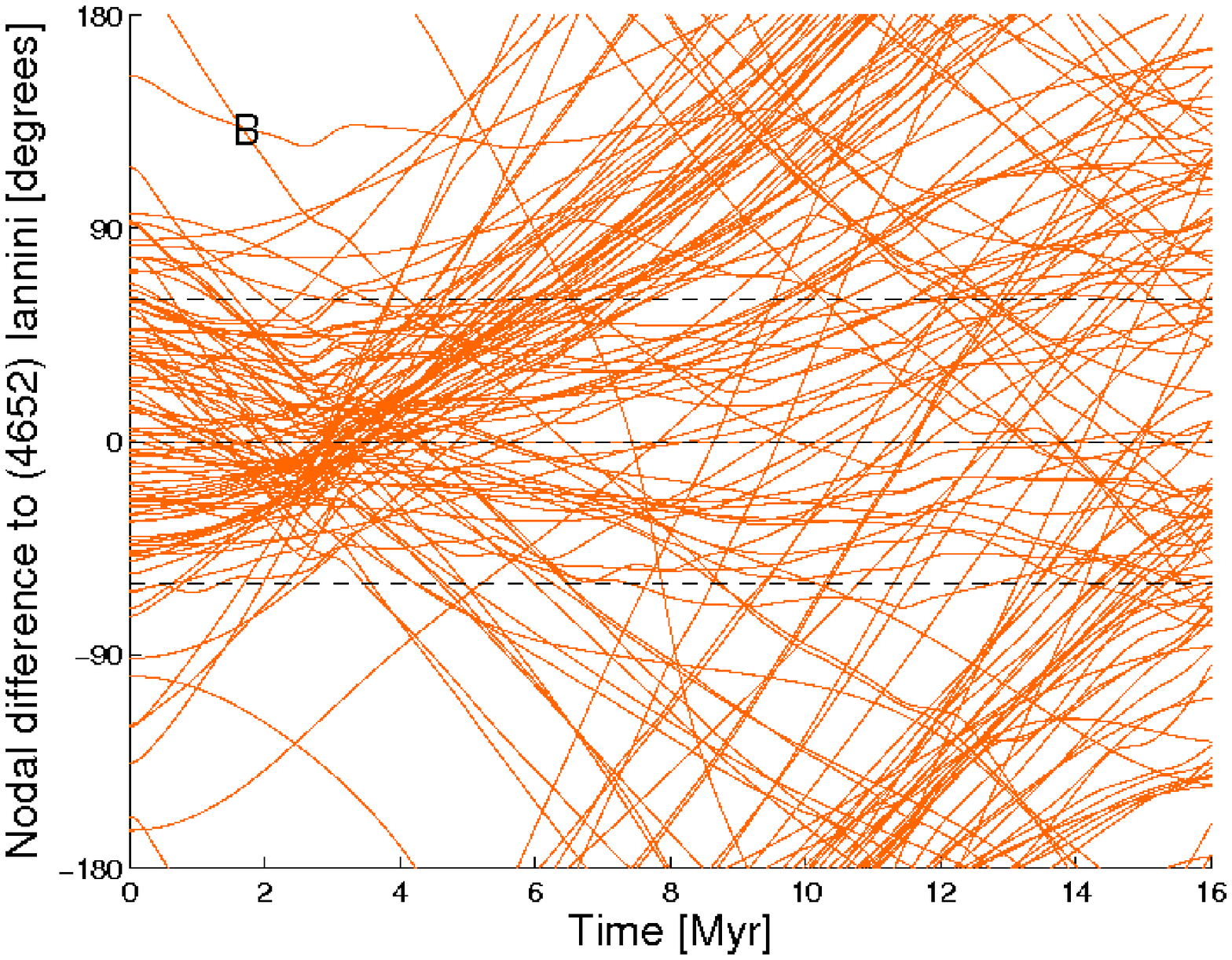}
 \end{minipage}
 \caption{The past evolution of the digitally filtered differences in
  pericenter (panel~A) and nodal (panel~B) longitudes for the $100$ largest
  Nele members with respect to the orbit of (4652) Iannini (the convergence
  pattern remains preserved if more asteroid orbits were shown but the
  figures would get saturated). Only the gravitational perturbations of
  planets were taken into account in the simulation. The horizontal lines
  display the $0^{\circ}$ and the $\pm 60^{\circ}$ levels for reference.}
\label{fig: Past_conv}
\end{figure*}

Following the approach described in \citet{Carruba_2016a}, we considered
the case of two reference orbits with respect to which difference 
$\Delta\varpi$ and $\Delta\Omega$ were computed: (i) (1547) Nele,
and (ii) (4652) Iannini, the two largest bodies in the family. In an
ideal world, which of the reference body is selected should not matter,
the choice of the largest fragment is typically justified by expecting
the least influence by the non-gravitational effects. In our case,
this would be (1547) Nele. However, we have noted above that the 
pericenter longitude of this body is slightly offset from the values of
other large members in the family (see Fig.~\ref{fig: sec_angles_dist}).
The reason is not known, but it may be caused by either stochastic effects
of close encounters to the most massive bodies in the main belt and/or
an asymmetric ejection velocity field with which the largest fragments
were launched. Such a slight discrepancy is not observed for 
(4652) Iannini, the second largest asteroid in the family. This
justifies our second choice for the reference orbit. While both result
in the same conclusion, we show in Fig.~\ref{fig: Past_conv} the
case of (4652) Iannini as a reference orbit. This is because the
data is visually more straightforward to interpret.

To obtain the age estimate for the Nele family, we first computed a
${\chi}^2$-like variable of the form:
\begin{equation}
 {\chi}_{0}^{2} = \sum_{i=2}^{N_{\rm ast}}({\Delta {\varpi}_{i}}^2
 +{\Delta {\Omega}_{i}}^2),
 \label{eq: chi2_age}
\end{equation}

\noindent where $N_{\rm ast}$ is the number of family members, 342, in our case.
At this moment, ${\chi}_{0}^{2}$ is plainly a sum of secular angle
differences to the reference asteroid and thus has dimension rad$^2$.
In our method, the best estimate of the family age corresponds to a minimum in 
these angular separations $\Delta {\varpi}_{i}$ and $\Delta {\Omega}_{i}$, or 
a minimum of the target function given in
Eq.~(\ref{eq: chi2_age}). In this case, we use (1547) Nele as a reference
body, thus ${\Delta {\varpi}}_i={\varpi}_{i}-{\varpi}_{\rm Nele}$ and
${\Delta {\Omega}}_i={\Omega}_{i}-{\Omega}_{\rm Nele}$. We output our data
every $10^3$~yr, but to prevent perturbing effects of a high-frequency
noise in ${\chi}_{0}^{2}$ values, both differences were filtered with a
low-pass digital Fourier filter. This procedure helped to
remove all frequency terms with periods shorter than $10^{5}$~yr 
\citep[e.g.,][]{Carruba_2010}. We tried other reasonable period
thresholds in our filtering procedure, but all resulted in very similar
conclusions. We also verified that the same results were obtained when
the orbit of (4652) Iannini was taken as a reference instead of (1547)
Nele. 

Once a minimum value of the ${\chi}_{0}^2$-like variable is obtained,
we estimated the statistical mismatch $\sigma$ of the best-fit 
convergence solution taking
$\Delta {\varpi}_{i}\simeq \Delta {\Omega}_{i}\simeq \sigma$.
With the definition (\ref{eq: chi2_age}) we have

\begin{equation}
 {\sigma}^2=\frac{({{\chi}_{0}}^2)_{\rm min}}{N_{\rm df}},
  \label{eq: chi2_err}
\end{equation}

\noindent where $N_{\rm df} = 2\,N_{\rm ast}-3$ is the number of degrees of
freedom. Note there were $2\,(N_{\rm ast}-1)$
number of pairs contributing by angular differences $\Delta {\varpi}_{i}$
and $\Delta {\Omega}_{i}$ with respect to the reference orbit, and an
additional factor $1$ is due to fitting the age of the family. In
quantitative terms, we have $({{\chi}_{0}}^2)_{\rm min} = 270.3$~rad$^2$ 
reached at the $3.6$~Myr epoch, and thence we obtain
${\sigma}^2 = 0.397$~rad$^2$. This corresponds to $\sigma\simeq 36.0^\circ$,
certainly quite more than the level at which fragments in the Nele family were
initially dispersed in either of the secular angles (i.e., a fraction of
a degree\footnote{The value of $\sigma$ may reduce if we
    eliminate from our sample the asteroids with the largest
    values of Lyapunov characteristic exponent ($LCE$), as also performed
    by other authors for the Veritas and Theobalda asteroid families
    \citep{Nesvorny_2003, Novakovic_2010}.  Since, however, in our sample
    there are just 14 asteroids with $LCE$ larger than that of (1547) Nele
    itself, and since the effect of removing them is to reduce $\sigma$ by
    just a degree, to $\simeq 34.9^\circ$, we preferred in this work to work
    with the completed Nele family sample.}).
At a plain face, we would need to reject the solution.
However, we know that the model used is only approximate since it does
not take into account fine details such as the thermal accelerations or
effects of additional massive bodies in the main asteroid belt. As a result,
we shall assume that the reached convergence level of $\simeq 36^\circ$
approximately corresponds to the possible limit of our crude model.
With this simple approach, we shall adopt that the obtained best-value
dispersion $\sigma$ corresponds to the uncertainty of the data points.

\begin{figure}
 \centering
 \centering \includegraphics [width=0.45\textwidth]{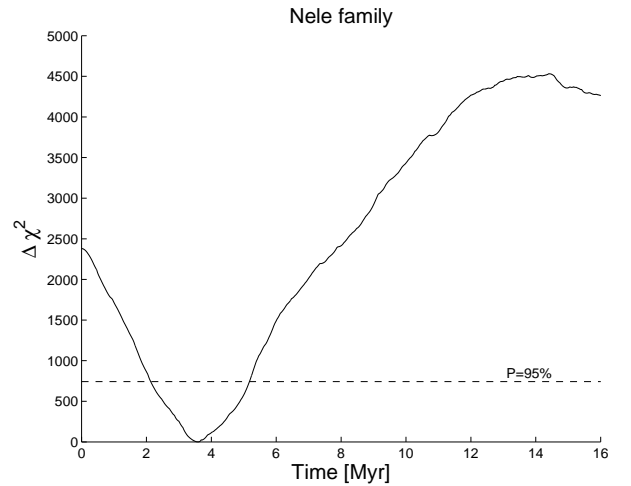}
 \caption{The time behavior of the $\Delta {\chi}^2 =
  {\chi}^2-{\chi}^2_{\rm min}$ variable defined in Eq.~(\ref{eq: new_chi2})
  for differences in the secular angles with respect to (1547) Nele,
  filtered with a cutoff of 10$^5$ yrs. The dashed line shows the
  95\% probability confidence level defined by $\Delta {\chi}^2\simeq
  742.8$. The age of the Nele family thus appears to be $3.6\pm 1.6$~Myr.}
\label{fig: nele_chi}
\end{figure}

This procedure allows us to construct a proper, now non-dimensional $\chi^2$
variable using the relationship

\begin{equation}
 {\chi}^{2} = \sum_{i=2}^{N_{\rm ast}}\left[\left(\frac{{\Delta {\varpi}_{i}}}{{\sigma}}
  \right)^2 +\left(\frac{\Delta {{\Omega}_{i}}}{{\sigma}}\right)^2\right].
 \label{eq: new_chi2}
\end{equation}

\noindent
With this target function defined, we formally use the least-squares
statistics to determine the single fitted parameter and its formal uncertainty, 
namely the age of the family. To that end we construct
$\Delta {\chi}^2 = {\chi}^2-{\chi}^2_{\rm min}$, where
${\chi}^2_{\rm min}\simeq 1$. Obviously, the formal best-fit solution of Nele
age still corresponds to the epoch where $\chi^2={\chi}^2_{\rm min}\simeq 1$.
Confidence level of this solution is set by some choice of $\Delta {\chi}^2$
value. For instance, the often used 95\% confidence level is obtained for
$\Delta {\chi}^2\simeq 742.8$. This is because our solution has $N_{\rm df}=
2\,N_{\rm ast}-3 = 679$ degrees of freedom, and the 95\% confidence level is
obtained from the cumulative distribution function of a ${\chi}^2$ variable,
expressed as:

\begin{equation}
 F(x,N_{\rm df}) = \frac{\gamma(\frac{N_{\rm df}}{2},\frac{x}{2})}{
  \Gamma(\frac{N_{\rm df}}{2})},
 \label{eq: chi2_cum}
\end{equation}

\noindent where $x=0.95$ stands for the chosen confidence level, $\gamma
(\frac{N_{\rm df}}{2},\frac{x}{2})$ is the lower incomplete gamma function,
and $\Gamma(\frac{N_{\rm df}}{2})$ is the gamma function. This analysis 
results in a Nele family
age of $3.6\pm1.6$~Myr. Note that this value is in good agreement 
with the value obtained in Sec.~4 (see Fig.~\ref{fig: std_nele}, panel~A)
from considerations of the Nele family dispersal in secular angles forward
in time.  Fig.~\ref{fig: nele_chi} shows the time dependence of
$\Delta {\chi}^2$ and determination of the 95\% confidence level interval.

Our simulation described above, in which we included only the gravitational
perturbations from planets, proved the tendency of
Nele-family orbits to converge within the past Myrs in both secular angles.
Still, the best attained level of $\simeq 36^\circ$ at which nodes and 
pericenters statistically approached the reference orbit of (1547) Nele or
(4652) Iannini is quite large. Recall that the initial conditions at the
family formation should produce a scatter of only a fraction of a degree.
Obviously, the effects of orbital chaos and additional gravitational
perturbers, such as massive asteroids, imply our simulation was not perfect and,
presumably, when these are included the convergence would be improved.
Additionally, previous studies demonstrated that especially for small
members in the families, those with $D\leq 2-3$~km, one needs also to account
for the thermal accelerations known as the Yarkovsky effect
\citep[e.g.,][]{Bottke_2002,Vokrouhlicky_2015}. \citet{Nesvorny_2004}
have shown that the Yarkovsky effect does not primarily perturb the secular
angles of the orbits in a direct way. Instead, the influence is indirect by
secularly
changing the orbital semimajor axis $a$. This is because the precession rate
of these angular variables due to the planetary perturbations sensitively
depends on the orbital semimajor axis, and therefore the Yarkovsky 
change in $a$ propagates into the $\Omega$ and $\varpi$ values. In what
follows we thus aim at implementing this improvement, and we try to
refine the orbital convergence of Nele members by accounting for the
Yarkovsky effect in their orbital evolution.

The Yarkovsky effect depends on a number of physical parameters, including
the spin state and surface thermal inertia, which are not known for
virtually all Nele members. The only information we were able to find in
the literature is the pole orientation for (1547) Nele that implies
its obliquity to be $50^\circ\pm 15^{\circ}$ \citep{Durech_2016}. We shall use
this constraint, but for all other asteroids in the family we must account
for a range of possible Yarkovsky effect values. This way for each body
we consider a certain number of Yarkovsky clones \citep[for more details
see, e.g.,][]{Carruba_2016a, Carruba_2017}. In particular, we estimate
the maximum Yarkovsky drift-rate $da/dt$ for (1547) Nele to be
$1.5\times 10^{-11}$ au~yr$^{-1}$ \citep[e.g.,][]{Bottke_2002,Vokrouhlicky_2015}.
Because the diurnal variant of the Yarkovsky effect typically dominates
the seasonal, the $\simeq 50^\circ$ obliquity of this body implies
a drift-rate of $1.5\times 10^{-11}\,\cos 50^\circ \simeq 9.6\times 10^{-12}$ 
au~yr$^{-1}$. Note that, when applied for the backward integration the sign
of $da/dt$ has to be reversed in our simulation. Each other body in the
Nele family was represented by $71$ Yarkovsky clones with $da/dt$ values
uniformly sampling a range between minimum and maximum estimated drift-rates
scaled from that of (1547) Nele, i.e., following the $da/dt \propto 1/D$
dependence on size $D$ \citep[e.g.,][]{Bottke_2002,Vokrouhlicky_2015}.

The whole sample of 24212 Yarkovsky clones (one for 1547~Nele) were numerically
integrated backward in time over a $21$~Myr interval using
${\tt SWIFT\_RMVS3\_DA}$. 
This code derives from the ${\tt SWIFT\_RMVS3}$ package in the swift-family and
has additionally an implementation of the Yarkovsky effect as described
in \citet{Nesvorny_2004}. We again used a $10$~day timestep and output data
every $10^3$~yr. Post-processing Fourier filtering techniques, similar to
those mentioned above, allowed us to remove high-frequency noise in the angular
differences $\Delta\varpi$ and $\Delta\Omega$ of all clones with respect to the
reference orbit of (1547) Nele. 

Out of the 71 clones of each particle, we then selected the one that showed
the best convergence to the reference orbit of (1547) Nele at each timestep. 
Using the target function in Eq.~(\ref{eq: chi2_age}), we first obtained 
an estimate of the minimum $({{\chi}_{0}}^2)_{\rm min} = 17.27$~rad$^2$
at $5.7$~Myr. We then repeated the method outlined above, namely assigned
a formal uncertainty $\sigma$ of the datapoints.  For the case of the
simulation that includes the Yarkovsky effect the number of degrees
of freedom is now $N_{\rm df} = 2(N_{\rm ast}-1) -(N_{\rm ast}-1) -1 = N_{\rm ast}-2$,
since there were  $2\,(N_{\rm ast}-1)$ number of pairs contributing by
angular differences, one estimated parameter (the family age) and we selected
the best Yarkovsky clone for $N_{\rm ast}-1$ family members (formally
fitting the relevant $da/dt$ values). Using Eq.~(\ref{eq: chi2_err}) with 
this new value of $N_{\rm df} =340$ we obtained ${\sigma}^2\simeq 0.050$~rad$^2$,
or $\sigma\simeq 12.9^\circ$. This is now a much better convergence solution
than before, obviously the result of the Yarkovsky effect contribution to 
the secular angles evolution (especially for smaller members in the family).
Note, however, that the $\sigma$ value is still at least an order of
magnitude larger than expected from the initial dispersion of secular
angles in the Nele family. The remaining mismatch is likely due to the
gravitational effects of massive bodies in the main belt and overall
chaotic dynamics in this zone of orbital phase space. It is also possible,
that some of the Yarkovsky drift-rates $da/dt$, assigned to the clones
in our second simulation, were overestimated to compensate these 
remaining dynamical effects that were not included in our model.

\begin{figure}
 \centering
 \centering \includegraphics [width=0.45\textwidth]{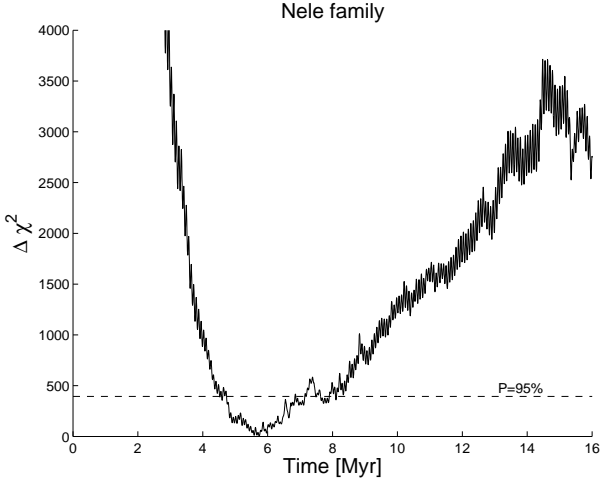}
 \caption{The time behavior of $\Delta{\chi}^2={\chi}^2-{\chi}^2_{\rm min}$
  variable defined in Eq.~(\ref{eq: new_chi2}) for a dynamical model
  where the thermal accelerations were taken into account. At each
  timestep, $10^3$~yr, we selected the optimum Yarkovsky drift-rate
  $da/dt$ for each Nele members (except for (1547) Nele, which has a
  constant value) in order to minimize $\chi^2$ value. The formal minimum
  with $\chi^2_{\rm min}\simeq 1$ occurs for $5.7$~Myr epoch. The
  95\% probability confidence level is shown by the dashed
  horizontal line.}
\label{fig: Nele_age}
\end{figure}

Applying now the best-fit $\sigma$ value to (\ref{eq: new_chi2}), 
a non-dimensional $\chi^2$ target function, we can again formally follow
the least-square method to determine the fitted parameters. 
As mentioned above, at each timestep at which results were output from
our simulation (i.e., $10^3$~yr), we selected the best possible $da/dt$
value of each of the Nele asteroids to obtain minimum $\chi_0^2$ or
$\chi^2$ value. This way we separate the solution of the family age from
the solution of the Yarkovsky drift-rates, withdrawing the possibility of
proper correlations of these many parameters. As above, we formally construct
$\Delta {\chi}^2 = {\chi}^2-{\chi}^2_{\rm min}$ using Eq.~(\ref{eq: new_chi2})
as a function of time (see Fig.~\ref{fig: Nele_age}). For $N_{\rm df} = 
340$ the 95\% confidence level corresponds to $\Delta {\chi}^2\simeq 394.6$
(Eq.~\ref{eq: chi2_cum}).

As seen in Fig.~\ref{fig: Nele_age}, at this confidence level we would 
conclude that the Nele family is $5.7^{+1.5}_{-1.3}$~Myr, which barely 
overlaps with the time interval found in the conservative simulation
(i.e., $3.6\pm1.6$~Myr). Recall also that an age older than $4.5$~Myr seems 
unlikely in view of the current clustering of $\varpi$ angles for the 
largest members ($H < 16$) of the Nele family, as discussed in
Sect.~\ref{sec: p_age_est}. Therefore the difference in the formal
age values obtained with the conservative simulation and, especially,
the one including the Yarkovsky effect requires some consideration.
One possibility of the discrepancy, advocated below, is due to
an asymmetric distribution of the Yarkovsky drift rates of the Nele 
family.

\begin{figure}
\centering
\centering \includegraphics [width=0.45\textwidth]{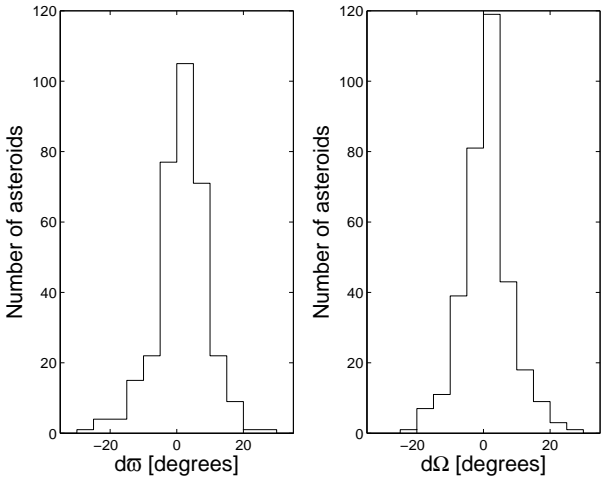}
\caption{Distributions of $\Delta {\varpi}$ and $\Delta {\Omega}$ of the
 best Yarkovsky clones at the Nele family formal age in our solution with
 the thermal accelerations included. Standard deviation of both variables
 is $\simeq 7.3^\circ$, indicating a significant improvement compared to the
 solution without the thermal effects.}
\label{fig: dom_dvarpi}
\end{figure}

Before we go to this speculative issue, we first show the distribution of 
$\Delta {\varpi}_i$ and $\Delta {\Omega}_i$ values for the formally 
best-fitting solution at $5.7$~Myr epoch in the past (the differences are 
with respect to the orbit of (1547) Nele; Fig.~\ref{fig: dom_dvarpi}).
The standard deviation of both distributions is small, namely
$7.3^{\circ}$ and $7.2^{\circ}$, which is at the above estimated level
$\sigma\simeq 12.9^\circ$. This confirms the ability of the dynamical
model to result in a much better convergence level than the purely
gravitational model above. Interestingly, the initial asymmetry of
the longitude of perihelion of (1547) Nele with respect to the
largest asteroids in its family has been well absorbed in the solution
(note the mean value of $\Delta {\varpi}_i$ on the left panel of
Fig.~\ref{fig: dom_dvarpi} is fairly close to zero).

\begin{figure}
 \centering
 \centering \includegraphics [width=0.45\textwidth]{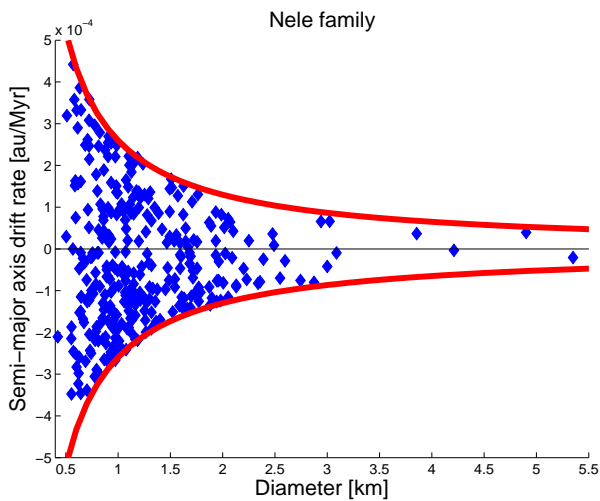}
 \caption{Semimajor axis drift-rates $da/dt$ for $321$ Nele family members
  with $D\leq 5.5$~km. These values correspond to the best-fit
  convergence of the family at $5.7$~Myr ago (see Fig.~\ref{fig: Nele_age}).
  Red lines show the estimated limiting values from the linearized theory of
  the Yarkovsky effect.}
\label{fig: drift_da}
\end{figure}

Next, we analysed the semimajor axis drift-rates $da/dt$ for the
Yarkovsky clones selected at the best-fit age solution at $5.7$~Myr 
ago. Fig.~\ref{fig: drift_da} shows their values as a function of
the body's size $D$. By definition, the $da/dt$ values for the clones 
are bound in the minimum/maximum interval estimated from the linear 
Yarkovsky theory \citep[e.g.][]{Bottke_2002,Vokrouhlicky_2015}.
The fact that these values do not pile-up towards the extreme values
is a good sanity check. Instead, they are roughly uniformly
distributed within allowed limits with: (i) a preference toward 
negative $da/dt$ values ($\simeq 61$\% of the sample), and (ii) a 
slight deficit values near $da/dt\simeq 0$ for small Nele members 
($D\leq 1.5$~km, say). The first, may be explained by a small contribution 
of the seasonal variant of the Yarkovsky effect as noted in the case 
of the Veritas family by \citet{Carruba_2017}. However, it could also 
result from an initial anisotropy of the rotation poles of Nele members 
having, for instance, preferentially obliquities larger than $90^\circ$.
The second feature in Fig.~\ref{fig: drift_da}, notably lack of 
$da/dt\simeq 0$ values for small Nele members, could be
indicative of the YORP effect tilting obliquities of the asteroids
toward extreme values as first detected in the case of the Karin
family by \citet{Carruba_2016a}. Note, however, that these results
should be taken with a reserve. This is because a proper analysis
of the uncertainties of the determined $da/dt$ values would likely 
indicate they basically span the whole available interval of values.

As a further check, we also determined the distribution of $da/dt$ values for 
younger ages of the Nele family. For example, when it is assumed that the
Nele family is $3.6$~Myr old (the formal best solution in a model where
only the gravitational perturbations were taken into account), the 
distribution of $da/dt$ becomes skewed toward
negative values with many asteroids having values near the maximum
drift rate permitted by the Yarkovsky theory. Such a non-random
distribution would be unusual. This is probably related to the fact
that Nele itself is offset in both angles from the rest of the family
at the present time (see Fig.~\ref{fig: sec_angles_dist}). This observation
may, though, explain the difference in the formal Nele ages in our
two models discussed in this Section. Assume, for instance, that the
true initial distribution of obliquities in the Nele family was
skewed towards values larger than $90^\circ$, with some $da/dt$ values
even exceeding our conservative limits. Our model with the thermal
accelerations included would not find these solutions near the
nominal age in the gravitational-only model. Rather, the formal
age would be pushed to older values where convergence would be 
achieved at the expense of a more symmetric, but false, set of the 
Yarkovsky clones. Without additional information, however, we cannot
decide which of the cases is correct and we do not have
an a priori reason to reject formal solutions with the Nele age
slightly exceeding $5-6$~Myr.

\section{Conclusions}
\label{sec: conc}

Our results could be summarized as follows:

\begin{itemize}

\item We revisited the Nele family membership by using the most
 up-to-date catalog of proper orbital elements, updating thus earlier
 findings in \citet{Milani_2014} and \citet{Nesvorny_2015}. The Nele family
 is a very compact cluster of asteroids in a rather isolated zone of the
 central main belt. No other important asteroid clusters were found in
 Nele vicinity, making thus the analysis of this family easier. None of the
 linear or principal non-linear secular resonances play an important role
 in the dynamical evolution of the Nele group. Moderate level of chaoticity
 of orbits in the Nele family, revealed by Lyapunov characteristic 
 exponents, is likely related to unusual three-body resonances not 
 involving planet Jupiter, four-body resonances, or two-body resonances
 with Venus.
  
\item We also reviewed the physical properties of asteroids in the Nele
 region, finding the available data are still rather scarce. Only $7$
 objects have taxonomic information from SDSS, and only $14$ have WISE,
 NEOWISE, AKARI or IRAS albedo data. Based on this limited information,
 we find that the Nele family is an S-type family, confirming previous
 analysis. The probable largest fragment of its namesake family, namely
 asteroid (1547) Nele, however shows some irregularity: (i) its spectrum
 has been preliminarily classified as TD in Tholen's taxonomic scheme,
 and (ii) its geometric albedo was found $0.20$, significantly lower
 then for other members in the cluster. It is possible, though, that both
 these data are uncertain, and more observations of this asteroid will
 be needed to clarify this issue.

\item The Nele family is rare among its class by indicating statistical
 clustering of the longitude of perihelion and node in orbits of the
 largest members \citep[e.g.,][]{Nesvorny_2003}. This property indicates
 its youth. Forward integration of a synthetic Nele family suggests
 the observed clustering of the perihelia is consistent with an age
 smaller then $4.5$~Myr. The age may be also estimated by the V-shape
 method discussed in \citet{Spoto_2015}, notably by interpreting slope of
 the delimiting lines of the family in the $(a,1/D)$ plane of parameters
 \citep[the equivalent method in the $(a,H)$ plane was introduced by][]
 {Vokrouhlicky_2006a}. This would result in an age of $\simeq 15$~Myr,
 neglecting the effect of initial fragment dispersal. On a flip
 side of the same coin, one could neglect the orbital evolution of the
 family fragments by the Yarkovsky effect and use the $(a,1/D)$
 data to estimate the characteristic speed at which the fragments in the
 Nele family were initially ejected from the parent body. This would
 lead to $\simeq 17$ m~s$^{-1}$. The reality is a combination of the two
 effects. Observing that the estimated escape velocity from (1547) Nele
 is $\simeq 10$ m~s$^{-1}$, the likely partition is $2:1$ in favor of the
 initial velocity effect. This would consistently come back to the
 rough age estimate of $\simeq 5$~Myr.
  
\item Inspired by previous success in the case of Karin and Veritas
 families \citep[and also a set of very small and young families
 reported, e.g., in][]{Nesvorny_2006} we attempted to improve the
 age estimate for the Nele family by numerical propagation of its
 orbits backward in time. The goal of this approach is to achieve
 as close convergence of the orbits as possible, in our case
 monitored by behavior of the secular angles only. We conducted two
 such simulations: (i) first, including only the gravitational
 perturbations of the planets, and (ii) second, including additionally
 the dynamical effects of the thermal accelerations. Each time we
 numerically integrated the orbits of $342$ members in the Nele family
 backward in time for $20$~Myr. The first set of runs, with
 gravitational perturbations included, lead to a poorer convergence.
 At best the perihelia and nodes statistically approached the orbit
 of (1547) Nele to about $36^\circ$. Assuming this is a reasonable limit
 of the simple model, we concluded that the Nele family is
 $3.6\pm1.6$~Myr old. As expected, the second model where thermal
 accelerations were modeled for each of the $342$ asteroids in the family
 lead to significant improvement of the convergence. At best, our solutions
 achieved $\simeq 7.5^\circ$ convergence level of perihelii and nodes.
 The nominal age found with the integration including Yarkovsky drift
 rates is $5.7^{+1.5}_{-1.3}$~Myr, barely compatible with the
 gravitational-only solution above. We suspect that the age difference
 in our two models may be due to a preferentially
 retrograde spin state of many fragments in this family.
 
\end{itemize}

Our analysis confirmed the very young age of the Nele family. This
implies that it remains the best source candidate of the J/K dust band
at $\simeq 12^\circ$ proper inclination \citep[e.g.,][]{Nesvorny_2003}.
An improvement of the Nele age estimate may require apriori constraints
on some, or all, of the free parameters in the model. For instance,
if we were to know obliquity values for some Nele members, the solution
will likely be quite more deterministic. At first sight this
may look like science fiction, however, we note the enormous improvements
in pole solutions for many main belt asteroids \citep[e.g.,][]{Durech_2015}.
Assuming that high-quality sparse photometry from ground- and space-based
upcoming surveys will become available, we may expect pole solutions
for thousands (if not more) of main belt objects.

\section*{Acknowledgments}

We are grateful to Andrea Milani and to the reviewer of this paper,
Bojan Novakovi\'{c}, for comments and suggestions that helped
to improve the quality of this paper. We would like to thank the S\~{a}o
Paulo State Science Foundation (FAPESP) that supported this work via the
grants 16/04476-8 and 2013/15357-1, and the Brazilian National Research Council
(CNPq, grant 305453/2011-4). DV's work was funded by the Czech Science
Foundation through the grant GA18-06083S. DN's work was supported by the NASA
SSW program. VC was a visiting scientist at the Southwest Research Institute
while some of the research for this work was developed. We acknowledge the 
use of data from the Asteroid Dynamics Site (AstDyS)
(http://hamilton.dm.unipi.it/astdys, \citet{Knezevic_2003}).
This publication makes also use of data products from the Wide-field 
Infrared Survey Explorer (WISE) and Near-Earth Objects (NEOWISE), which
are a joint project of the University of California, Los Angeles, and the
Jet Propulsion Laboratory/California Institute of Technology, funded by the
National Aeronautics and Space Administration.

\bsp

\label{lastpage}

\end{document}